\documentclass[cqg, groupedaddress, twocolumn]{revtex4}
\usepackage{graphicx,epsf,amssymb,amsmath,latexsym,epsfig}

\begin{document}

\title{Stabilization of Neutral Thin Shells By Gravitational Effects From Electric Fields}

\vspace{.3in}

\author{Eduardo I. Guendelman}
\email{guendel@bgu.ac.il}
\author{Idan Shilon}
\email{silon@bgu.ac.il}

\affiliation{Physics Department, Ben-Gurion University of the Negev, Beer-Sheva 84105, Israel}

\vskip.3in

\begin{abstract}

We study the properties of a system consisting of an \textit{uncharged} spherically symmetric two dimensional extended object which encloses a stationary point charge placed in the shell's center. We show that there can be a static and stable configuration for the neutral shell, using only the gravitational field of the charged source as a stabilizing mechanism. In particular, two types of shells are studied: a dust shell and a string gas shell. The dynamical possibilities are also analyzed, including the possibility of child universe creation.

\end{abstract}

\maketitle

\setcounter{equation}{0}

\section{Introduction}

The formation of global or local topological defects, and in particular domain walls, is known to happen during a spontaneous symmetry breaking phase transitions, when a nontrivial topology exists. In the 
cosmological context, these objects are likely to occur in the very early universe and therefore might have important 
implications for the creation and evolution of our universe \cite{vil, khlop}.

Spherical solutions of domain walls are usually unstable towards collapse. However, prior studies have shown that some modifications of the effective surface tension (e.g radially dependent surface tension) \cite{ahr, port}, as well as considering charged bubbles \cite{kuch, gog}, can yield a static and stable shell solution. However, this kind of stabilization takes place even before considering gravitational effects. Stable configurations of domain walls can give rise to many interesting physical models (e.g models for elementary particles). In this work we wish to explore a different way of achieving stabilization of a two dimensional extended object using the gravitational effects of a massive charged source on an \textit{uncharged} spherical shell. Hence, the stabilization in the case presented here is induced purely from gravitational effects.

In order to analyze the problem we use a very well-known and studied classical
system, known as a \textit{general relativistic shell} \cite{isr}.
General relativistic shells, or bubbles, provide a non-trivial, gravitational system, whose
classical dynamics can be described by an intuitive set of equations with a
clear geometrical meaning. many useful analytical results are known and numerical methods
have also been employed.
General relativistic shells are a proper framework to describe vacuum bubble
formation and dynamics where the region over which the transition between the interior and exterior
of the shell is very small compared to the other scales of the
problem. This approximation is usually known as the thin wall approximation, which can also be expressed
in terms of the difference between the vacuum expectation values of a scalar field,
which is different in the two regions of spacetime.
Naturally, the thin layer over which the transition is realized can be
modeled by a mathematical surface which, on the physical side,
is equipped by an effective surface tension. 

In this paper, we wish to show that obtaining a static and stable domain walls configuration is indeed possible, even
without introducing additional mass terms. The major result of our analysis is that the mere presence of a global electric field (that is, an electric field which is present on both regions of the global spacetime manifold) can serve as a stabilizing 
mechanism for a vacuum bubble. Here, the source of this electric field is a massive point charge, located at the center of the vacuum bubble.

This paper is organized as follows: In section \ref{EMprop} we briefly analyze the general electromagnetic properties of the system. In section \ref{eom} we obtain the equation that determines the dynamics of a bubble that carries an arbitrary matter content on its surface. We then analyze this equations for two analytical cases: a bubble which carries dust on its surface (i.e a dust shell) and a bubble that carries a string gas like content. We show that dynamics of the bubble in both cases have similar properties and, in particular, that both cases can yield a stable and static bubble configuration. In section \ref{conc} we summarize our conclusions and give prospects for future studies.

\section{Analysis of the Electro-Magnetic Properties}
\label{EMprop}

Since we have a massive and stationary point charge located at the center of the bubble, the general metrics for both regions  of our set-up are Reissner-Nordstrom metrics. As the shell does not carry any charge, the charge parameter that appears in both metrics should be equal, leading to

\begin{equation}
\label{metric}
	\begin{array}{cc}
	ds^{2}_{-} = -\left(1 - \frac{2Gm}{r} + \frac{GQ^{2}}{r^{2}}\right) dt^{2} +  \vspace{4pt}\\ +
	\left(1 - \frac{2Gm}{r} +\frac{GQ^{2}}{r^{2}}\right)^{-1}dr^{2} + r^{2}d\Omega^{2}, & r < R,  \vspace{12pt}\\
	ds^{2}_{+} = -\left(1 - \frac{2GM}{r} + \frac{GQ^{2}}{r^{2}}\right) dt^{2} +  \vspace{4pt}\\ +
	\left(1 - \frac{2GM}{r} + \frac{GQ^{2}}{r^{2}}\right)^{-1}dr^{2} + r^{2}d\Omega^{2}, & r > R,
	\end{array}	
\end{equation}

where $G$ is Newton's constant, $m$ is the mass of the electric source, with charge $Q$, and $M$ is the mass-energy of the system as viewed by an external observer. The only non-zero component of the electromagnetic tensor will be $F_{0r} = - F_{r0}$. Therefore, Maxwell's equations reduce to the single equation

\begin{equation}
\label{ }
	\nabla_{r}F^{0r} = \frac{1}{\sqrt{-g}}\partial_{r}\left(\sqrt{-g}g^{00}g^{rr}F_{0r}\right) = 0,
\end{equation}

for $r \neq 0$. Thus, from classical electrodynamics, we have (for each region) $(\sqrt{g}g^{00}g^{rr})^{i}F_{0r}^{i} \propto Q$, where $i$ indexes the different regions. From the latter we find that the electric field is continuous across the shell 

\begin{equation}
\label{eps }
	F_{0r} = \frac{Q}{r^{2}}, ~ \forall ~ r.
\end{equation}

Finally, we find the energy density of the electric field to be

\begin{equation}
\label{eng1}
	T^{0}_{~0} = 
	-\frac{Q^{2}}{2r^{4}}, ~ \forall ~ r.
\end{equation}

\section{Dynamical Analysis}
\label{eom}

Now we turn to derive the effective potential which governs the bubble dynamics and look for at-least one local minimum point which fulfills the condition $V_{eff}(R_{min}) = 0$. From Israel's junction conditions \cite{isr}, the bubble equation of motion is

\begin{equation}
\label{mot}
	\epsilon_{-}\sqrt{A_{-}(R) +\dot{R}^{2}} - \epsilon_{+}\sqrt{A_{+}(R) +\dot{R}^{2}} = \frac{Gm(R)}{R} = \kappa(R),
\end{equation}

where the $\pm$ signs indicate the exterior and the interior regions of the shell, respectively, $A_{\pm}(R) = -g^{\pm}_{tt}$, where we consider only spherically symmetric metrics that satisfy the relation $g_{rr}g_{tt} = -1$ (as in Eq. (\ref{metric})), $m(R) = 4\pi R^{2} \sigma(R)$ describes the energy-matter content which is located on the surface of the bubble and the coefficients $\epsilon_{\pm} = \mbox{sgn}(n^{\mu}\partial_{\mu}r)|_{\mathcal{M}_{\pm}}$ determine if the radial coordinate $r$ is decreasing ($\epsilon_{\pm} = -1$) or increasing ($\epsilon_{\pm} = +1$) along the normal coordinate to the brane, $n^{\mu}$. The global manifold that we are considering here will not contain any horizons. Thus, both coefficients, $\epsilon_{\pm}$, will be positive. A more thorough discussion about this point will be given below.

Let us write the equation of state for the matter on the bubble surface as $p=\omega\sigma$, where $p$ is the two dimensional isotropic pressure, $\omega$ is a constant and $\sigma$ is the surface energy density. Then, from conservation of energy

\begin{equation}
\label{ener}
	\sigma(R) = \sigma_{0}R^{-2(1+\omega)}
\end{equation}

where $\sigma_{0}$ is a constant. For example, $\omega = -1$ represents a constant surface tension and $\omega = - 1/2$ represents a matter that can be interpreted as a gas of strings living on the surface of the bubble \cite{kolb, gas}. Putting Eq. (\ref{ener}) into the equation of motion (\ref{mot}) we get

\begin{equation}
\label{ }
	\kappa(R) = 4\pi G\sigma_{0}R^{-(1 + 2\omega)} = \kappa_{0}R^{-(1 + 2\omega)},
\end{equation}

where $\kappa_{0} = 4\pi G\sigma_{0}$. Equation (\ref{mot}) is equivalent to the equation of motion of a particle moving in one dimension under the influence of an effective potential \cite{blau}, of the form $\dot{R}^{2} + V_{eff} = 0$. In our case, the effective equation

\begin{equation}
\label{ }
	\begin{array}{c}
	\dot{R}^{2} + 1 - \frac{\kappa_{0}^{2}R^{-2(2\omega +1)}}{4} - \frac{G(M+m)}{R} + \vspace{4pt}\\ + \frac{GQ^{2}}		{R^{2}} - \frac{G^{2}(M-m)^{2}}{\kappa_{0}^{2}}R^{4\omega}  = 0,
	\end{array}
\end{equation}

defines the effective potential to be

\begin{equation}
\label{ }
	\begin{array}{c}
	V_{eff}(R) = 1 - \frac{\kappa_{0}^{2}R^{-2(2\omega +1)}}{4} - \frac{G(M+m)}{R} + \vspace{4pt}\\ + \frac{GQ^{2}}		{R^{2}} -\frac{G^{2}(M-m)^{2}}{\kappa_{0}^{2}}R^{4\omega}.
	\end{array}
\end{equation}

Let us now concentrate on two specific cases that have a clear physical meaning: $\omega = -1/2$ and $\omega = 0$. For these two values of the equation of state parameter the effective potential is tractable and can be solved in closed form.

It is known that a gas of strings in $n$ spatial dimensions satisfies the equation of state $p = -\sigma/n$. Therefore, the first case describes a gas of strings which is located on the surface of the bubble, while the second choice corresponds to dust living on the wall.

We wish to show now that in both cases the potential can have a local minimum which satisfies $V_{eff}(R_{min})=0$, after some fine tuning. Let us begin with the stringy gas bubble.

\subsection{Stringy Gas Bubble}
\label{SGB}

The effective potential for this kind of bubble is ($\omega = -1/2$)

\begin{equation}
\label{ }
	V_{eff}^{S.G}(R) = 1 - \frac{\kappa_{0}^{2}}{4} - \frac{G(M+m)}{R} + 			\frac{GQ^{2}}{R^{2}} -\frac{G^{2}(M-m)^{2}}{\kappa_{0}^{2}R^{2}}.
\end{equation}

Solving the equation $V_{eff}^{S.G}(R) =0$, we are lead to a polynomial equation of rank 2. By imposing the demands that the discriminant will equal zero and that 
$\kappa_{0}^{2}<4$, we ensure that we will have a global minimum which satisfies $V_{eff}(R_{min}) = 0$.

The discriminant of the effective potential function is given by

\begin{equation}
\label{ }
	\Delta = G^{2}(M+m)^{2} + (\kappa_{0}^{2} - 4)\left(GQ^{2}-G^{2}\frac{(M-m)^{2}}{\kappa_{0}^{2}}\right)
\end{equation}

and by demanding $\Delta = 0$ we arrive at the condition:

\begin{equation}
\label{q1}
	Q^{2} = \frac{G(M+m)^{2}}{4 - \kappa_{0}^{2}} + \frac{G(M-m)^{2}}{\kappa_{0}^{2}}.
\end{equation}

The radius of curvature of the stable bubble is thus given by

\begin{equation}
\label{ }
	R_{min} = 2G\frac{M+m}{4-\kappa_{0}^{2}},
\end{equation}

with $R_{min}$ being positive under the imposed conditions. 

Now we turn to ask if $R_{min}$ is located behind any horizons. In principle, there might be two different horizons in each region. But, if $Q^{2}$ is bigger than $GM^{2}$, then there will be no horizons in the system (remembering that $M\geq m$, since we do not consider negative surface tension). Comparing Eq. \ref{q1} to $GM^{2}$ is equivalent to comparing the function $f_{1}(x)$ and the quantity $g_{1}$, where:

\begin{eqnarray}
f_{1}(x) & = & x^{2} + x (\kappa_{0}^{2}-2) + 1 \\
g_{1} & = &  \frac{\kappa_{0}^{2}}{4}(4-\kappa_{0}^{2}) 
\end{eqnarray}

with $x=m/M$ being the ratio between the masses (which is constrained $0<x<1$).

So the condition $Q^{2}<GM^{2}$ is now $f_{1}(x) < g_{1}$. The minimum value of $f_{1}(x)$ is located at $x_{min} = (2 - \kappa_{0}^{2})/2$, when $\kappa_{0}^{2} < 2$, and at $x=0$ when $\kappa_{0}^{2}>2$. Looking at the minimum value of $f_{1}$ we find that it satisfies $f_{1}(x_{min}) = g_{1}$. This means that $f_{1}(x)$ will always be greater than $g_{1}$ (i.e $Q^{2} > GM^{2}$) except for the case when the ratio between the masses is fine tuned to equal precisely $x_{min}$ and the surface tension is small enough (i.e  $\kappa_{0}^{2} < 2$). This limiting case means that there would be two degenerate horizons (i.e an extremal black hole like object). When $x\neq x_{min}$, we have a naked singularity at $r=0$, since this corresponds to $Q^{2}>GM^{2}$. When $m=(1- \kappa_{0}^{2}/2)M$ (i.e $x = x_{min}$), the bubble would sit exactly on the degenerate horizon (i.e $R_{min} = GM$) and therefore it would be a light-like brane. However, we consider here only a timelike motion for the brane and this particular case is merely a miscellaneous limiting situation.

The above result is actually an example of the more general relation between the effective potential and the metric coefficients which states that $A_{\pm}(R) - V_{eff}(R) \geq 0$. Thus, when the effective potential is definite positive, as in the cases studied here, the geometries $\mathcal{M}_{\pm}$ can not contain any horizons. A proof for this relation is given in Ref. \cite{ste}.

\begin{figure}[htp]
	\begin{center}
	\includegraphics[scale=0.42]{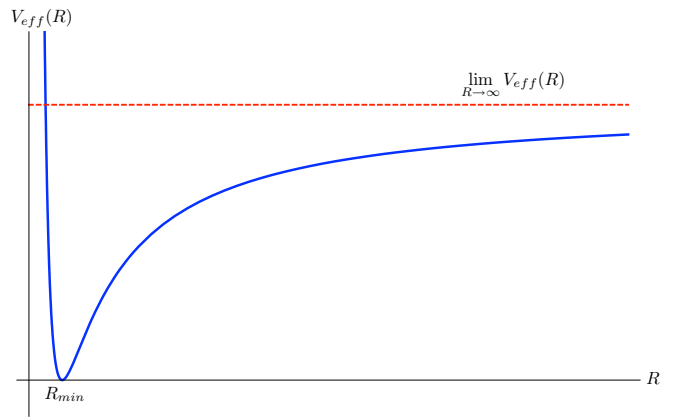}
	\end{center}
	\caption{The general characteristics of $V_{eff}(R)$ (continuos line). This graph corresponds to both cases analyzed here: $\omega = -1/2 ~\mbox{and}~ \omega = 0$, which show similar characteristics of the effective potential. The dashed line corresponds to the limit of $V_{eff}(R)$ at $R \to \infty$: $1- \kappa_{0}^{2}/4$ in the $\omega = -1/2$ case and $1 - \left(G(M-m)/\kappa_{0}\right)^{2}$ in the $\omega = 0$ case.} \label{fig:figure1}
\end{figure}

\subsection{Dust Shell}

For the case of $\omega = 0$, the effective potential for this case is

\begin{equation}
\label{ }
	V_{eff}^{DUST}(R) = 1 - \frac{\kappa_{0}^{2}}{4R^{2}} - \frac{G(M+m)}{R} + 	\frac{GQ^{2}}{R^{2}} -\frac{G^{2}(M-m)^{2}}{\kappa_{0}^{2}}.
\end{equation}

In the same manner of section \ref{SGB}, by solving the equation $V_{eff}^{DUST}(R) =0$ we encounter again a simple quadratic equation. The conditions for having a stable solution now are (note the different units of $\kappa_{0}$ in this case) $G^{2}(M-m)^{2} < \kappa_{0}^{2}< 4GQ^{2}$ and 

\begin{equation}
\label{ }
	\Delta =  G^{2}(M+m)^{2}  + (\kappa_{0}^{2} - 4GQ^{2})\left(1 - \frac{G^{2}(M-m)^{2}}{\kappa_{0}^{2}}\right) 	= 0.
\end{equation}
 
The latter can, again, be written as a constraint on the charge:

\begin{equation}
\label{ }
	Q^{2} = \frac{\kappa_{0}^{2}}{4G}\frac{4G^{2}Mm+\kappa_{0}^{2}}{\kappa_{0}^{2}-G^{2}(M-m)^{2}}
\end{equation}

The radius of curvature for the stable bubble is now given by

\begin{equation}
\label{ }
	R_{min} = \frac{1}{2}\frac{G(M+m)\kappa_{0}^{2}}{\kappa_{0}^{2}-G^{2}(M-m)^{2}},
\end{equation}

which is, again, positive under the imposed conditions. 

Looking for the location of $R_{min}$ relative to the horizons, we continue in analogy to the prior case, where now

\begin{eqnarray}
f_{2}(x) & = & G^{3}M^{4}(1-x)^{2} + GM^{2}\kappa_{0}^{2}x \\
g_{2} & = &\kappa_{0}^{2}\left(GM^{2} - \frac{\kappa_{0}^{2}}{4G}\right).
\end{eqnarray}

$f_{2}(x)$ has one minimum at $x_{min}=1 - \frac{\kappa_{0}^{2}}{2(GM)^{2}}$, when $\kappa_{0}^{2} < 2(GM)^{2}$, and at $x=0$ when $\kappa_{0}^{2}>2(GM)^{2}$. It can be easily verified that $f_{2}(x_{min}) = g_{2}$. The conclusions are the same as before: there will be no horizons unless $\kappa_{0}^{2} < 2(GM)^{2}$ and $x=x_{min}$, where the latter means that the two horizons are degenerate and located at $R = GM$, exactly where the bubble would sit. Again, this demonstrates the more general relation $A_{\pm}(R) - V_{eff}(R) \geq 0$ \cite{ste}.

\subsection{Dynamical Solutions}
\label{exp}

Here we would like to address the question: what happens to the two types of bubbles we have considered earlier, if the conditions for time independence are not satisfied? Obviously, one option is that no dynamics is possible (when $\Delta < 0$). The other option is that the effective potential will be negative in some region, which will allow a kinetic energy for the shell. Here we have two cases: a finite or an infinite region where $V_{eff} < 0$. The first case (finite region) corresponds to bounded solutions, or a 'breathing' bubble \cite{sakai}, where the bubble starts off at some radius, expands to a maximum radius value and then shrinks back to the initial radius. On the other hand, the second case corresponds to a bubble which begins with a finite radius and then expands to an infinite size. 

The question is where to the bubble expands? In other words, does it blow up and 'eat' the surrounding manifold or could it have a way of expanding without replacing its surroundings? \cite{ansref}. Peculiar as it may sound, due to the structure of the spacetime manifold, the second option is indeed possible. When wormholes are present, the bubble can make its way and expand to infinity by creating its own space completely disconnected from the original spacetime. This solution is usually referred to a 'child universe'.

In order to determine if wormholes might be present in the solution we need to calculate the extrinsic curvature tensor induced on the shell. This was done, of course, in Eq. (\ref{mot}). For spherically symmetric metrics the relevant component of the extrinsic curvature is $K_{\theta\theta} \propto \partial g_{\theta\theta}/\partial n$ so that the coefficients $\epsilon_{\pm}$ determine the behavior of the radial coordinate in the direction orthogonal to the trajectory of the bubble, hence allowing us to determine the existence of wormholes. Calculating the signs of these coefficients for the two types of bubbles we are considering, we find that for the string gas bubble

\begin{eqnarray}
\mbox{sgn}(\epsilon_{+}) & = & \mbox{sgn}\left(\frac{2G(M-m)}{R} - \kappa_{0}^{2}\right), \\
\mbox{sgn}(\epsilon_{-}) & = & +1
\end{eqnarray}

and therefore the extrinsic curvature induced on the shell from the exterior changes sign at larger $R$ values. This is a generic characteristic for the presence of wormholes since when $\epsilon_{+}$ is negative the normal coordinate to the brane, which points from the bubble interior to its exterior, is pointing along a direction for which the radial coordinate is actually decreasing rather than, as in the more familiar possibility of non wormhole geometry, increasing. Thus, we conclude that there is a possibility for the bubble to expand to infinity disconnected from the original spacetime (i.e a child universe solution \cite{ansref}).

For the dust shell case, the trajectory constants signs read

\begin{eqnarray}
\mbox{sgn}(\epsilon_{+}) & = & \mbox{sgn}\left(\frac{2G(M-m)}{R} - \frac{\kappa_{0}^{2}}{R^{2}}\right), \\
\mbox{sgn}(\epsilon_{-}) & = & +1,
\end{eqnarray}

so in the limit $R \to \infty$ we see that $\epsilon_{+}$ is positive, indicating that the normal to the brane does not point to a direction in which the radial coordinate decreases. Hence, there is no child universe formation \cite{ansref}.

These issues will be studied in more detail in a future publication.

\section{Conclusions}
\label{conc}

To conclude, the electric charge causes a gravitational repulsive effect which balances the natural tendency of two dimensional extended objects to collapse and thus yields a static and stable shell configuration, even though the latter carries zero charge and does not interact directly with electric fields. 

We note that in the limiting case where the charge parameter of the interior and exterior solutions is zero our results coincide with those presented by Kijowski, Magli and Malafarina \cite{magli}, where they reviewed the dynamics of spherical timelike shells by matching two different Schwarzschild spacetimes and also analyzing the canonical formulation of such systems. 

In a future research we will study the semi-classical quantization of the bounded excitations of the string gas shell, where the structure of the effective potential , which contains a flat region as $r \to \infty$ (independent of the mass), we see that there is the possibility of 'ionization' which could be responsible for a dynamical creation of a universe, since for bigger radii the solutions will approach those studied in references \cite{ans1, ans}, which represent child universe creation.   

In contrast, for the case of a dust shell, the ionization does not produce a child universe, but instead it is simply an 'expansion' where the dust shell achieves the critical 'escape velocity' necessary to expand to infinity in the existing space, i.e not creating a new space of its own (a child universe).

\vskip.3in

\centerline{{\bf Acknowledgments}} 
We wish to thank Stefano Ansoldi for helpful discussions and useful suggestions and comments on the manuscript. The research of IS was partially supported by The Israel Science Foundation grant no 470/06 and by the Israeli 'Commercial and Industrial Club'.

\vskip.4in


\vskip1.6in


\begin{thebibliography}{99}

\bibitem{vil}
A. Vilenkin and E.P.S. Shellard, \textit{Cosmic Strings and Other Topological Defects} (Cambridge University Press, Cambridge,1994).

\bibitem{khlop}
M.Y. Khlopov, \textit{Cosmoparticle Physics}, (World Scientific, 1997).

\bibitem{ahr}
A. Davidson and E.I. Guendelman, Phys. Lett. {\bf B 251}, 250 (1990).

\bibitem{port}
E.I. Guendelman and J. Portnoy, Class. Quant. Grav. {\bf 16}, 3315 (1999); arXiv: gr-qc/9901066.

\bibitem{kuch}
K. Kuchar, Czech. J. Phys. B {\bf 18}, 435 (1968).

\bibitem{gog}
A.T. Barnaveli and M.Ya. Gogberashvili, Theor. and Math. Phys. {\bf 113}, 2 (1997).

\bibitem{isr}
W. Israel, Nouvo Cimento {\bf 44B}, 1 (1966); {\bf 48B}, 463(E) (1967).

\bibitem{kolb}
E.W. Kolb, Astroph. J. {\bf 344}, 543 (1989).

\bibitem{gas}
S. Alexander, R.H. Brandenberger and D. Easson, Phys. Rev. D {\bf 62}, 103509 (2000).

\bibitem{blau}
S.K. Blau, E.I. Guendelman and A.H. Guth, Phys. Rev. D {\bf 35}, 1747 (1987).

\bibitem{ste}
S. Ansoldi, PoS(QG-Ph), 004  (2007); arXiv: 0709.2741 [gr-qc].

\bibitem{sakai}
For another example of breathing bubble see: E.I. Guendelman and N. Sakai, Phys. Rev. D {\bf 77}, 125002 (2008); arXiv: 0803.0268 [gr-qc].

\bibitem{ansref}
For a recent review: S. Ansoldi, E.I. Guendelman and I. Shilon, in proceedings of Conference on Black Holes and Naked Singularities, Milan, Italy (2007), arXiv: 0711.2198 [gr-qc].

\bibitem{magli} 
J. Kijowski, G. Magli ang D. Malafarina, Gen. Rel. Grav. {\bf 38}, 1697 (2006).

\bibitem{ans1}
S. Ansoldi and E.I. Guendelman, Prog. Theor. Phys. {\bf 120}, 985 (2007); arXiv: 0706.1233 [gr-qc].

\bibitem{ans}
S. Ansoldi, E.I. Guendelman and I. Shilon, \textit{in preparation}.

\end{thebibliography}
\end{document}